\documentclass[10pt,twocolumn,english,aps,pra]{revtex4-2}
\usepackage[T1]{fontenc}
\usepackage[utf8]{inputenc}
\usepackage{babel}
\usepackage{amsmath}
\usepackage{amsthm}
\usepackage{graphicx}
\usepackage[unicode=true,pdfusetitle,
 bookmarks=true,bookmarksnumbered=false,bookmarksopen=false,
 breaklinks=false,pdfborder={0 0 1},backref=false,colorlinks=false]
 {hyperref}
\hypersetup{
 colorlinks=true}

\makeatletter
\numberwithin{equation}{section}
\numberwithin{figure}{section}

\usepackage{units}
\usepackage{babel}
\usepackage{float}
\usepackage{graphicx}
\usepackage{babel}
\usepackage{enumitem}
\renewcommand{\theenumi}{\alph{enumi}}
\usepackage{tikz}
\usepackage{vmargin}
\setmargins{1.5cm}       
{0.5cm}                        
{17.5cm}                      
{25.42cm}                    
{14pt}                           
{1cm}                           
{0pt}                             
{0cm}                           
\usepackage{amsmath}
\usepackage{amssymb}
\usepackage{cancel}
\usepackage[nodayofweek,yyyymmdd]{datetime}
\usepackage{mdframed}
\definecolor{green}{RGB}{0,128,0}
\addto\captionsenglish{}

\makeatother

\begin{document}
\global\long\def\blue#1{\textcolor{blue}{#1}}%
\global\long\def\red#1{\textcolor{red}{#1}}%
\global\long\def\green#1{\textcolor{green}{#1}}%
\global\long\def\purple#1{\textcolor{purple}{#1}}%
\global\long\def\orange#1{\textcolor{orange}{#1}}%

\global\long\def\it#1{\textit{\textrm{#1}}}%
\global\long\def\un#1{\underline{\textrm{#1}}}%
\global\long\def\br#1{\left( #1 \right)}%
\global\long\def\sqbr#1{\left[ #1 \right]}%
\global\long\def\curbr#1{\left\{  #1 \right\}  }%
\global\long\def\braket#1{\langle#1 \rangle}%
\global\long\def\bra#1{\langle#1 \vert}%
\global\long\def\ket#1{\vert#1 \rangle}%
\global\long\def\abs#1{\left|#1\right|}%
\global\long\def\mb#1{\mathbf{#1}}%
\global\long\def\doublebraket#1{\langle\langle#1 \rangle\rangle}%

\title{Vibropolaritonic Reaction Rates in the Collective Strong Coupling
Regime: Pollak-Grabert-Hänggi Theory}
\author{Matthew Du}
\thanks{Both authors contributed equally to this work.}
\author{Yong Rui Poh}
\thanks{Both authors contributed equally to this work.}
\author{Joel Yuen-Zhou}
\email{joelyuen@ucsd.edu}

\affiliation{\emph{Department of Chemistry and Biochemistry, University of California
San Diego, La Jolla, California 92093, USA}}
\date{January 6, 2023}
\begin{abstract}
Following experimental evidence that vibrational polaritons, formed
from collective vibrational strong coupling (VSC) in optical microcavities,
can modify ground-state reaction rates, a spate of theoretical explanations
relying on cavity-induced frictions has been proposed through the
Pollak-Grabert-Hänggi (PGH) theory, which goes beyond transition state
theory (TST). However, by considering only a single reacting molecule
coupled to light, these works do not capture the ensemble effects
present in experiments. Moreover, the relevant light-matter coupling
should have been $\sqrt{N}$ times smaller than those used by preceding
works, where $N\approx10^{6}-10^{12}$ is the ensemble size. In this
work, we explain why this distinction is significant and can nullify
effects from these cavity-induced frictions. By analytically extending
the cavity PGH model to realistic values of $N$, we show how this
model succumbs to the polariton ``large $N$ problem'', that is,
the situation whereby the single reacting molecule feels only a tiny
$1/N$ part of the collective light-matter interaction intensity,
where $N$ is large.
\end{abstract}
\maketitle

\section*{Introduction}

Vibrational strong coupling (VSC) occurs when molecular vibrational
modes interact strongly with infrared photon modes, typically confined
in an optical cavity, thus forming new light-matter hybrid modes known
as vibrational polaritons \citep{Long2015,Shalabney2015}. With microcavities
such as Fabry-Pérot cavities, these interactions are only appreciable
in the presence of a macroscopic number of molecules, that is, VSC
is a collective effect \citep{CamposGonzalezAngulo2020,MartinezMartinez2019}.
Over the past decade, vibrational polaritons formed from microcavities
have been experimentally shown to influence (1) ground-state chemical
reactivities \citep{Thomas2016,Lather2019,Hirai2020,GarciaVidal2021,Thomas2019,Lather2021,Lather2022,Ahn2022}
and (2) vibrational energy transfer processes \citep{Dunkelberger2016,Xiang2018,Xiang2020,Pang2020},
creating a field known as vibropolaritonic chemistry. These experiments,
conducted under the following conditions: \citep{YuenZhou2021} $
\global\long\def\theenumi{\text{C}\arabic{enumi}}%
$
\begin{enumerate}
\item $N\approx10^{6}-10^{12}$ molecules \emph{collectively} coupled to
the cavity,
\item in the absence of optical pumping, that is, the reaction relies purely
on \emph{thermal fluctuations},
\end{enumerate}
reported the following observations: $
\global\long\def\theenumi{\text{O}\arabic{enumi}}%
$
\begin{enumerate}
\item The cavity may either \emph{enhance} or \emph{suppress} reaction rates;
\item Rate modification by the cavity is optimum when the cavity mode is
\emph{resonant} with the reactant, spectator and/or solvent vibrational
modes, and
\item occurs only for the cavity mode at normal incidence ($k=0$).
\end{enumerate}
$
\global\long\def\theenumi{\alph{enumi}}%
$Unfortunately, there remains a dearth of theoretical models that successfully
explain all five features. In particular, the first class of transitions
has, in the absence of VSC, been well-explained by thermal adiabatic
rate models \citep{Peters2017} such as transition state theory (TST).
Along this vein, pioneering studies have attempted to incorporate
VSC effects into a classical TST model \citep{Galego2019,CamposGonzalezAngulo2020},
only to find that the activation energy remains unchanged once the
often-neglected dipole self-energy term of the photon mode is included
\citep{Li2021} (although this conclusion has been contested by models
that account for vibrational quantum effects \citep{Yang2021}). The
transmission prefactor may, however, be reduced by VSC \citep{Li2021}
through a dynamical caging effect similar to the Grote-Hynes theory
\citep{Grote1980}, yet this result fails to account for the collective,
resonance and $k=0$ features present in experiments (i.e. features
C1, O2 and O3). This sparked a series of works that considered additional
effects such as anharmonicities \citep{Hernandez2019,Triana2020,CamposGonzalezAngulo2021,Ribeiro2018},
multiple cavity modes \citep{Hoffmann2020,Ribeiro2022}, inter-mode
energy redistributions \citep{Schafer2021,Wang2022-single,Wang2022-collective,Fischer2022},
and disorders \citep{Sidler2021,Botzung2020}. A summary of these
theoretical results will be presented in an upcoming perspective \citep{CamposGonzalezAngulo2022}.
Note that considerable efforts have also been devoted to the second
class of transitions \citep{CamposGonzalezAngulo2019,Vurgaftman2020,Phuc2020,Du2021,Du2022,Poh2022,Cao2022}
and will not be discussed here.

This paper joins the wave of some recent works \citep{Sun2022,Lindoy2022,Philbin2022}
that explored the possibility of cavity-induced frictions. Through
classical trajectory-based simulations over a range of bath frictions,
Sun and Vendrell reported cavity-mediated rate enhancements in the
low-friction regime that peak when the reacting vibrational mode is
resonant with the cavity mode \citep{Sun2022} (thereby fulfilling
observation O2). Their result is consistent with the Pollak-Grabert-Hänggi
(PGH) description \citep{Pollak1989}, an analytical adiabatic rate
model that also considers weak energy exchange between the system
and bath modes. Here, system refers to a reactive mode, so stronger
system-bath couplings (also known as frictions) allow the reacting
system to more easily acquire energy from the bath to cross the barrier,
thereby accelerating the reaction. Note that the PGH theory is a non-Markovian
generalisation of the Kramers turnover model, first predicted by Kramers
\citep{Kramers1940} and later solved by Mel'nikov and Meshkov \citep{Mel'nikov1986}.
Importantly, PGH theory goes beyond TST and is unlike the Grote-Hynes
theory \citep{Grote1980}, which may be reduced to TST \citep{Pollak1986}.
In the context of VSC, Lindoy et al. interpreted the cavity mode as
an effective bath mode, which exerts cavity-induced friction on the
system \citep{Lindoy2022}. As such, the cavity accelerates chemical
reactions when its coupling to the reactive mode is stronger than
that of the molecule's inherent bath modes. In both aforementioned
works \citep{Sun2022,Lindoy2022}, the cavity was assumed to have
a single photon mode. Later, Philbin et al. extended the model to
an imperfect cavity with multiple confined modes \citep{Philbin2022}
and made qualitatively similar observations apart from sharper resonances
and weaker cavity effects.

\begin{figure}
\includegraphics[width=1\columnwidth]{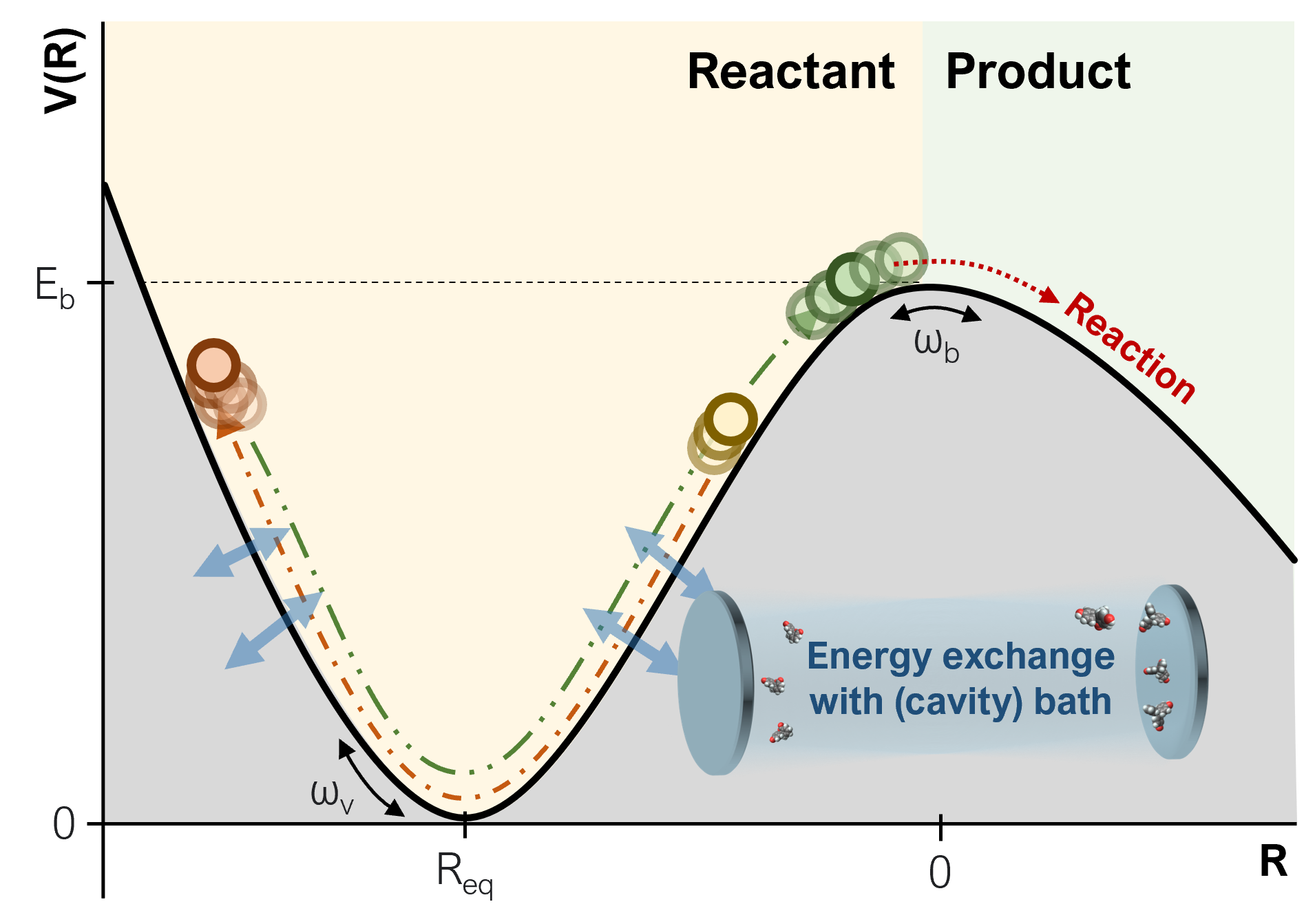}\caption{\label{fig:main}PGH theory and its application to VSC. The PGH particle
moves classically along a \textquotedblleft system\textquotedblright{}
reactive mode (coordinate $R$) coupled to a thermal bath. The potential
along the reactive mode has a barrier $E_{b}$ (position $R=0$, harmonic
frequency $\omega_{b}$), which separates a potential well (equilibrium
position $R=R_{\text{eq}}$, harmonic frequency $\omega_{v}$) from
the continuum; these two regions signify the reactant and product
regions respectively. The particle starts off with energy $E<E_{b}$
(yellow particle) and is unable to cross the barrier. Instead, it
moves to and fro between the barrier and the reactant region (red
and green paths), during which it exchanges energy with the thermal
bath through the system-bath couplings (or frictions). These couplings
are small so the particle may, at some point (green particle), acquire
sufficient energy from the thermal bath to cross the barrier ($E\ge E_{b}$)
and react. PGH theory estimates the rate of this process $-$ effectively,
it computes the rate of thermally-activated chemical reactions in
the low bath friction limit. When applied to VSC, the potential is
the adiabatic electronic ground-state potential energy surface. In
the single-molecule regime, the cavity mode serves as an effective
bath mode \citep{Lindoy2022} whereas, in the collective regime, the
cavity and additional $N-1$ non-reactive vibrational modes couple
to form two polariton modes, which then become effective bath modes.}
\end{figure}

These studies, while enlightening, were investigated for a single
(or few) molecule(s) interacting with light and are therefore not
fully representative of the ensemble effects observed in VSC involving
microcavities \citep{Thomas2016,Lather2019,Hirai2020,GarciaVidal2021,Thomas2019,Lather2021,Lather2022,Ahn2022}
(see condition C1). In addition, their numerical results were reported
using experimental values of light-matter interactions belonging to
an entire molecular ensemble, even though the single-molecule interaction
would have been more accurate. This distinction is highly non-trivial:
the appropriate single-molecule light-matter coupling $g$ is $\sqrt{N}$
times smaller than the experimentally-measured collective coupling
$g\sqrt{N}$, where $N$ is the number of confined molecules estimated
to be $\approx10^{6}-10^{12}$ \citep{Pino2015,Daskalakis2017}. Clearly
$g\not\approx g\sqrt{N}$. To demonstrate how collectivity changes
the effects of cavity-induced frictions, we analytically extend the
PGH model to include a macroscopic number of molecules $N$, each
of which interacts with a single cavity mode {[}Fig. (\ref{fig:main}){]}.
Using a $g\sqrt{N}$ value representative of experimental data, we
find that the purported cavity-mediated rate enhancements quickly
and expectedly vanish with increasing $N$ and are negligible for
realistic values of $N>20$. In particular, cavity-induced frictions
depend on $g$ to leading order in $g\sqrt{N-1}$, a result familiar
to the community of collective VSC \citep{Kansanen2022,Poh2022,Yang2021}.
Therefore, for a constant $g\sqrt{N-1}\approx g\sqrt{N}$, we find
that $g$ and thus cavity-mediated frictions diminish with realistic
values of $N$. This is a reminder of the polariton ``large $N$
problem'' \citep{MartinezMartinez2019}, that is, that any benefit
from the polaritons is often lost to the penalty of having a large
number ($N-1$) of non-reacting molecules compete with a single reacting
molecule for the cavity {[}Fig. (\ref{fig:coupling_orders}){]}. Finally,
we qualitatively argue why our observations remain valid even with
disorder and multiple cavity modes.

With reference to the five features described earlier, our model fully
addresses C1, C2 and O2 and partially addresses O1 by dealing only
with rate enhancements. Note that the collective effect described
by condition C1 is the novel part of this work.

\section*{Results and Discussions}

Outside the cavity, the PGH model \citep{Pollak1989} considers a
single reactive (system) mode of coordinate $R$ coupled to a harmonic
thermal bath of coordinates $\curbr{Q_{\alpha}}$ and frequencies
$\curbr{\omega_{\alpha}}$. The potential of the reactive mode has
a barrier of height $E_{b}$ separating a well from the continuum
{[}Fig. (\ref{fig:main}){]}. A particle (such as a single molecule)
moves along its reactive and bath modes classically; it starts from
a metastable state in the well and, through energy exchange with the
thermal bath modes, harnesses sufficient energy to cross the barrier,
signifying a reaction of which the rate may be computed. Working in
mass-weighted coordinates, the Hamiltonian for this model is
\begin{align}
H_{\text{PGH}} & =\frac{\dot{R}^{2}}{2}+V\br R\nonumber \\
 & \quad+\sum_{\alpha=1}^{\mathcal{N}}\sqbr{\frac{\dot{Q}_{\alpha}}{2}+\frac{1}{2}\br{\omega_{\alpha}Q_{\alpha}+\gamma_{\alpha}R}^{2}},\label{eq:H_PGH}
\end{align}
where $\curbr{\gamma_{\alpha}\in\mathbb{R}}$ are the system-bath
couplings (or frictions), and $V\br R$ is the potential along the
reactive coordinate $R$ and is modelled harmonically near the barrier
and well bottom (with imaginary frequency $i\omega_{b}$ and real
frequency $\omega_{v}$ respectively; $\omega_{b},\omega_{v}\in\mathbb{R}^{+}$).
It is convenient to work in the normal mode basis around the barrier
region, which comprises one unstable mode $u$ of imaginary frequency
$i\Omega_{b}$ ($\Omega_{b}\in\mathbb{R}^{+}$) and $\mathcal{N}$
stable modes $\curbr{s_{k}}$ of real frequencies $\curbr{\Omega_{k}\in\mathbb{R}^{+}}$.
This implies that (1) the reaction occurs along the unstable mode
since it has a barrier (of frequency $\Omega_{b}$), and (2) in the
weak system-bath coupling limit, the reactive mode is composed mostly
of the unstable mode such that contributions from the stable modes
may be used to characterise these system-bath couplings. Mathematically,
if we define $R=Q_{1}=\cdots=Q_{\mathcal{N}}=0$ when the particle
is at the barrier and expand $R$ in the stable-unstable mode basis
as
\begin{align}
R & =c_{00}u+\sum_{k=1}^{\mathcal{N}}c_{0k}s_{k}
\end{align}
with $c_{0k}\in\mathbb{R}$ for all $k=0,1,\cdots,\mathcal{N}$ and
$\sum_{k=0}^{\mathcal{N}}c_{0k}^{2}=1$, then the total system-bath
coupling may be characterised by
\begin{align}
\epsilon & =\sum_{k=1}^{\mathcal{N}}\frac{c_{0k}^{2}}{c_{00}^{2}}=\frac{1}{c_{00}^{2}}-1.\label{eq:epsilon}
\end{align}
PGH theory focuses on the weak system-bath coupling limit ($\epsilon\ll1$),
so we expect larger $\epsilon$ to improve energy exchange between
system and bath modes and therefore increase the reaction rate. Finally,
by analysing the energy flow between the stable and unstable modes
during the particle's path in the well (during which the stable and
unstable modes are no longer normal modes), the reaction rate is predicted
to be \citep{Pollak1989} 
\begin{align}
k & =\kappa k_{\text{1DTST}},\label{eq:k_PGH}
\end{align}
where $k_{\text{1DTST}}=\frac{\omega_{v}}{2\pi}e^{-\beta E_{b}}$
is the rate calculated using 1D transition state theory (1D-TST) and
is independent of system-bath couplings, and 
\begin{align}
\kappa & =\frac{\Omega_{b}}{\omega_{b}}\nonumber \\
 & \quad\times\exp\sqbr{\frac{1}{\pi}\int_{-\infty}^{\infty}\frac{dy}{1+y^{2}}\,\ln\br{1-e^{-\br{\beta\Delta E}\frac{1+y^{2}}{4}}}}\label{eq:kappa}
\end{align}
is the transmission factor due to system-bath couplings and serves
as a proxy measure of the bath's effects on the reaction rate. Here,
$\beta=\br{k_{\text{B}}T}^{-1}$, with $T$ as temperature. Also,
$\Delta E$ characterises the system-bath energy exchange and has
the form 
\begin{align}
\Delta E & =\frac{1}{2}\sum_{k=1}^{\mathcal{N}}\frac{c_{0k}^{2}}{c_{00}^{2}}\abs{\tilde{F}\br{\Omega_{k}}}^{2},\label{eq:DeltaE}
\end{align}
where $\tilde{F}\br{\Omega_{k}}$ is the Fourier transform of the
effective force experienced by the unstable mode. The latter may be
solved analytically for a piecewise differentiable parabolic potential
of the form
\begin{align}
V\br R & =\begin{cases}
\frac{1}{2}\omega_{v}^{2}\br{R-R_{\text{eq}}}^{2} & R\le R',\\
-\frac{1}{2}\omega_{b}^{2}R^{2}+E_{b} & R>R',
\end{cases}\label{eq:V(R)}
\end{align}
where $R_{\text{eq}}=-\sqrt{2E_{b}\br{1/\omega_{v}^{2}+1/\omega_{b}^{2}}}$
and $R'=R_{\text{eq}}\omega_{v}^{2}/\br{\omega_{v}^{2}+\omega_{b}^{2}}$
are derived from making both $V\br R$ and $\partial_{R}V\br R$ continuous
at $R=R'$ (thus giving two conditions that solve for two unknowns).
The result is
\begin{align}
\abs{\tilde{F}\br{\Omega_{k}}}^{2} & =\frac{4\br{R'/c_{00}}^{2}\Omega_{b}^{2}\br{\omega_{\text{eff}}^{2}+\Omega_{b}^{2}}^{2}}{\Omega_{k}^{2}\br{\Omega_{k}^{2}-\omega_{\text{eff}}^{2}}^{2}}\nonumber \\
 & \quad\times\sqbr{\Omega_{b}\sin\br{\Omega_{k}\tau/2}+\Omega_{k}\cos\br{\Omega_{k}\tau/2}}^{2},
\end{align}
where $\omega_{\text{eff}}=\sqrt{c_{00}^{2}\br{\omega_{v}^{2}+\omega_{b}^{2}}-\Omega_{b}^{2}}$
is the effective well frequency experienced by the unstable mode,
and $\tau$ is the system-bath interaction time obtained by solving
the coupled equations
\begin{align*}
 & \begin{cases}
\cos\omega_{\text{eff}}\tau=\br{\Omega_{b}^{2}-\omega_{\text{eff}}^{2}}/\br{\Omega_{b}^{2}+\omega_{\text{eff}}^{2}},\\
\sin\omega_{\text{eff}}\tau=-\br{2\Omega_{b}\omega_{\text{eff}}}/\br{\Omega_{b}^{2}+\omega_{\text{eff}}^{2}}.
\end{cases}
\end{align*}
Focusing on Eqs. (\ref{eq:k_PGH}), (\ref{eq:kappa}) and (\ref{eq:DeltaE}),
we find that stronger system-bath couplings, characterised by $\epsilon=\sum_{k=1}^{\mathcal{N}}c_{0k}^{2}/c_{00}^{2}$,
facilitate energy exchange $\Delta E$ between the modes and also
modify the unstable mode barrier frequency $\Omega_{b}$. While the
former increases $\kappa$, the reaction rate relative to that from
1D-TST, the latter may change $\kappa$ in either directions.

To apply PGH theory to VSC, Lindoy et al. considered the reactive
mode (with finite barrier like Eq. (\ref{eq:V(R)})) to be bilinearly
coupled to both the harmonic molecular bath and a single cavity mode
\citep{Lindoy2022}. This bilinear light-matter interaction originates
from the Pauli-Fierz nonrelativistic QED Hamiltonian \citep{Flick2017,Rokaj2018,Schaefer2020}
and may be interpreted as a cavity-induced friction in the PGH framework
(i.e. the cavity mode acts as an effective bath mode). Since the reactive
mode belongs to a single reacting molecule, its harmonic molecular
bath corresponds to its ``solvent'' environment. Chemical reactions
are rare events, and it is unlikely for two molecules to react simultaneously.
As such, to extend this model to the collective regime, we need to
include $N-1$ other non-reacting molecules, each with its own harmonic
(and, therefore, non-reactive) vibrational mode also coupled bilinearly
to the \emph{same} cavity mode and its own \emph{separate} harmonic
molecular bath. (Alternatively, this model may be interpreted as explicitly
describing the set-up of ``cooperative VSC'', whereby a small amount
of reactive species is placed in a sea of chemically inert molecules,
all of which couple to the cavity in the same fashion \citep{Lather2019,Lather2021,Wiesehan2021,Lather2022}.)

In the following paragraphs, we will show that, through a normal mode
transformation, we may rewrite the subsystem comprising the cavity
and $N-1$ non-reactive molecules into a pair of polariton modes bilinearly
coupled to the reactive mode. Just like the single-molecule models
\citep{Lindoy2022,Sun2022,Philbin2022}, these bilinear couplings
represent cavity-induced frictions, which accelerate the reaction
in the low-friction regime.

The Hamiltonian describing the above model is, in mass-weighted coordinates,
\begin{align}
H & =T_{\text{r}}+V_{\text{r}}\br R+H_{\text{nr-c}}+H_{\text{r-c}}+H_{\text{bath}},\label{eq:H-VSC}
\end{align}
where $T_{\text{r}}=\dot{R}^{2}/2$ is the kinetic energy of the reactive
mode with coordinate $R$, $V_{\text{r}}\br R$ is the piecewise differentiable
parabolic potential of the reactive mode as described by Eq. (\ref{eq:V(R)})
(well frequency $\omega_{v}$, barrier frequency $\omega_{b}$), 
\begin{align}
H_{\text{nr-c}} & =\sum_{j=1}^{N-1}\br{\frac{\dot{X}_{j}^{2}}{2}+\frac{\omega_{v}^{2}}{2}X_{j}^{2}}\nonumber \\
 & \quad+\frac{\dot{q}_{c}^{2}}{2}+\frac{1}{2}\br{\omega_{c}q_{c}+2g\sum_{j=1}^{N-1}X_{j}}^{2}\label{eq:H_nr-c}
\end{align}
represents the couplings between the cavity mode (coordinate $q_{c}$,
frequency $\omega_{c}$) and $N-1$ non-reactive vibrational modes
(coordinates $\curbr{X_{j}}$, frequency $\omega_{v}$), as well as
their kinetic and potential energies, 
\begin{align}
H_{\text{r-c}} & =\br{\omega_{c}q_{c}+2g\sum_{j=1}^{N-1}X_{j}}2gR+2g^{2}R^{2}\label{eq:H_r-c}
\end{align}
represents the couplings between the reactive mode and the subsystem
of cavity and $N-1$ non-reactive vibrational modes, and $H_{\text{bath}}$
represents all $N$ sets of molecular bath modes that belong to the
single reactive and $N-1$ non-reactive vibrational modes. As mentioned
earlier, this Hamiltonian may be derived from the cavity QED Hamiltonian
in the dipole gauge, under the cavity Born-Oppenheimer approximation,
after assuming all molecules to be in the adiabatic electronic ground
state \citep{Flick2017,Rokaj2018,Schaefer2020}; more details may
be found in Ref. \citep{Li2021}. Here, we consider only a single
cavity mode and assume that all $N-1$ non-reactive vibrational modes
have the same spatial alignments and frequencies $\omega_{v}$ as
the potential well of the reactive mode (the result remains qualitatively
unchanged under isotropic alignment of dipoles and will be explained
later). As such, all molecules couple equally to the cavity mode,
each with the same coupling amplitude of $g=-\boldsymbol{\mu}_{0}'\cdot\boldsymbol{\epsilon}/\sqrt{4\epsilon_{0}\mathcal{V}}$,
where $\boldsymbol{\epsilon}$ is the polarisation unit vector of
the cavity mode, $\mathcal{V}$ is the cavity's effective quantisation
volume, and $\boldsymbol{\mu}_{0}'$ is the linear change of the dipole
moment along each vibrational mode near its equilibrium position (well
bottom for the reactive mode), identical for all modes (i.e. $\boldsymbol{\mu}_{j}'=\boldsymbol{\mu}_{\text{r}}'=\boldsymbol{\mu}_{0}'$
for all $j=1,\cdots,N-1$). While $H_{\text{nr-c}}+H_{\text{r-c}}$
{[}Eqs. (\ref{eq:H_nr-c}) and (\ref{eq:H_r-c}){]}, which describe
couplings to the reactive mode, is not yet in the form of $H_{\text{PGH}}$
{[}Eq. (\ref{eq:H_PGH}){]}, a normal mode transformation will do
the trick. Exploiting the symmetries created by the assumptions above,
we can rewrite the $N-1$ degenerate non-reactive vibrational modes
into a single bright mode with coordinate
\begin{align}
Q_{\text{B}} & =\frac{1}{\sqrt{N-1}}\sum_{j=1}^{N-1}X_{j}
\end{align}
and $N-2$ dark modes with coordinates 
\begin{align*}
Q_{\zeta} & =\frac{1}{\sqrt{N-1}}\sum_{j=1}^{N-1}A_{j,\zeta}X_{j}, & \zeta & =2,\cdots,N-1,
\end{align*}
where the coefficients $\curbr{A_{j,\zeta}}$ are real-valued for
all $j$ and $\zeta$ and satisfy the orthonormality conditions of
\begin{align*}
\sum_{j=1}^{N-1}A_{j,\zeta} & =0 &  & \text{and} & \sum_{j=1}^{N-1}\frac{A_{j,\zeta}A_{j,\eta}}{N-1} & =\delta_{\zeta\eta},
\end{align*}
with $\delta_{\zeta\eta}$ representing the Kronecker delta. Then,
$H_{\text{nr-c}}$ {[}Eq. (\ref{eq:H_nr-c}){]} becomes
\begin{align}
H_{\text{nr-c}} & =\sum_{\zeta=2}^{N-1}\br{\frac{\dot{Q}_{\zeta}^{2}}{2}+\frac{\omega_{v}^{2}}{2}Q_{\zeta}^{2}}+\frac{\dot{Q}_{\text{B}}^{2}}{2}+\frac{\omega_{v}^{2}}{2}Q_{\text{B}}^{2}\nonumber \\
 & \quad+\frac{\dot{q}_{c}^{2}}{2}+\frac{1}{2}\br{\omega_{c}q_{c}+2g\sqrt{N-1}Q_{\text{B}}}^{2},\label{eq:H_nr-c-bright}
\end{align}
i.e. only the bright mode has the correct symmetry to couple with
the cavity. Performing a normal mode transformation on these two modes
gives two polariton modes with coordinates $Q_{\pm}$ and frequencies
$\omega_{\pm}$ (see Appendix 1). By expressing $H_{\text{nr-c}}$
{[}Eq. (\ref{eq:H_nr-c-bright}){]} and $H_{\text{r-c}}$ {[}Eq. (\ref{eq:H_r-c}){]}
in terms of the polariton modes, the Hamiltonian becomes
\begin{align}
H & =H_{\text{eff}}+H_{\text{bath}}+\sum_{\zeta=2}^{N-1}\br{\frac{\dot{Q}_{\zeta}^{2}}{2}+\frac{\omega_{v}^{2}}{2}Q_{\zeta}^{2}},\label{eq:H-VSC-new}
\end{align}
where
\begin{align}
H_{\text{eff}} & =\frac{\dot{R}^{2}}{2}+V_{\text{r}}\br R\nonumber \\
 & \quad+\sum_{\alpha=\pm}\sqbr{\frac{\dot{Q}_{\alpha}^{2}}{2}+\frac{1}{2}\br{\omega_{\alpha}Q_{\alpha}+2g_{\alpha}R}^{2}},\label{eq:H_eff}
\end{align}
such that the subsystem of cavity and $N-1$ non-reactive vibrational
modes forms a pair of effective polariton bath modes (coordinates
$Q_{\pm}$) that interact with the reactive mode through couplings
$g_{\pm}$. These system-polariton couplings $g_{\pm}$ are analogous
to the cavity-induced friction $g$ described by single-molecule models
\citep{Sun2022,Lindoy2022,Philbin2022} (for instance, compare Eq.
(\ref{eq:H_eff}) with Eq. (3) of Ref. \citep{Lindoy2022}, which
denotes $g$ by $\eta_{c}\sqrt{\omega_{c}/2}$). Note that the dark
modes (coordinates $Q_{2},\cdots,Q_{N-1}$) do not couple to the reactive
mode and may be neglected in our future analysis.

\begin{figure}
\includegraphics[width=1\columnwidth]{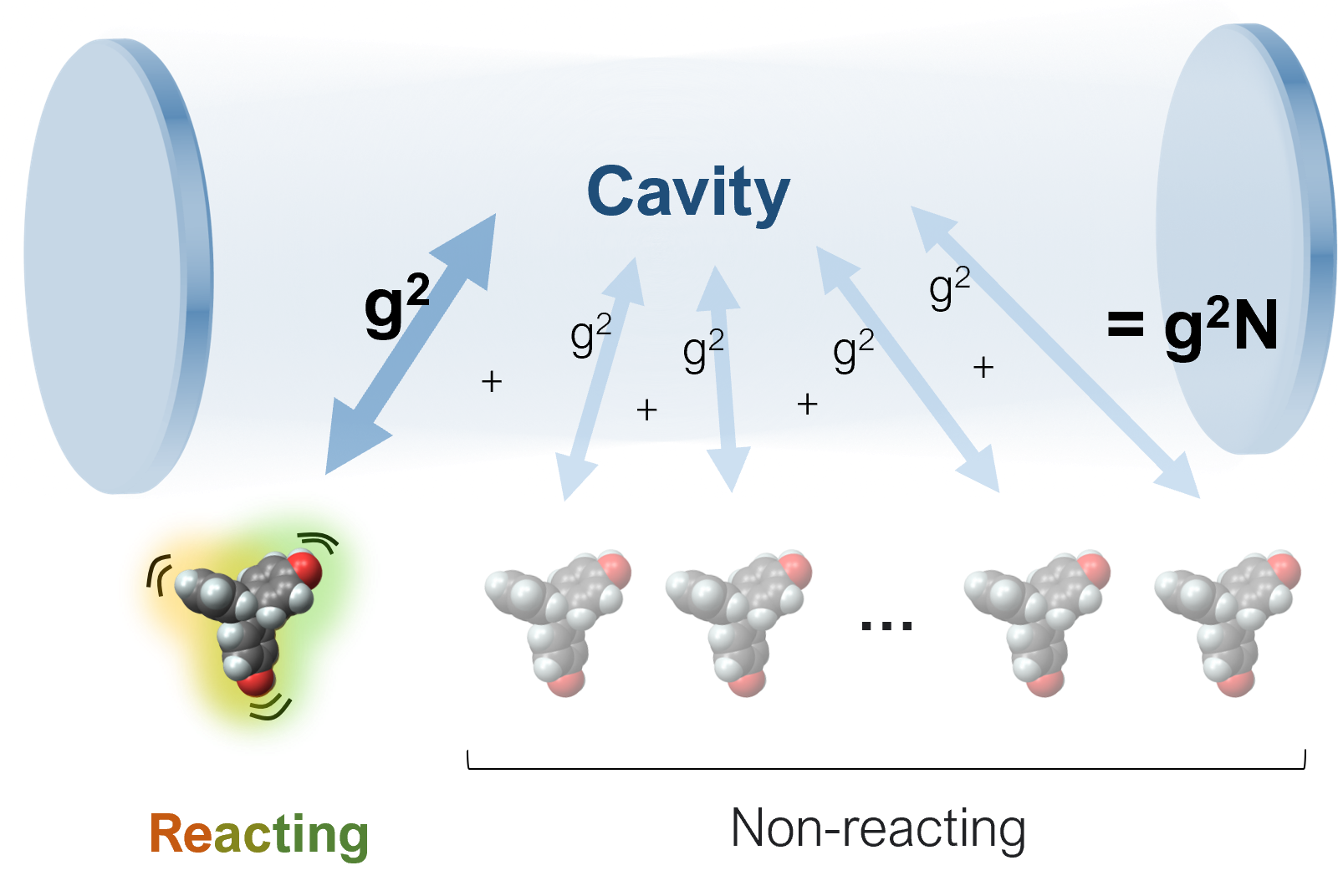}\caption{\label{fig:coupling_orders}Hierarchy of interactions with the reacting
molecule. The cavity couples collectively to all $N$ molecules with
intensity $g^{2}N$, so each molecule experiences an average coupling
intensity of $g^{2}$. From the reacting molecule's perspective, it
first couples directly and most strongly to the cavity with intensity
$g^{2}$. The same molecule also couples to the remaining $N-1$ non-reacting
molecules, but only through the cavity, thus making this a second-order
process with intensities proportional to the product of the two sub-processes'
couplings: $g^{2}$ and $g^{2}\protect\br{N-1}$. This effect is captured
in the series expansion of system-polariton couplings $g_{\pm}$ {[}Eq.
(\ref{eq:g_pm-analytical}){]} and is an important part of the polariton
\textquotedblleft large $N$ problem\textquotedblright .}
\end{figure}

\begin{figure}
\includegraphics[width=1\columnwidth]{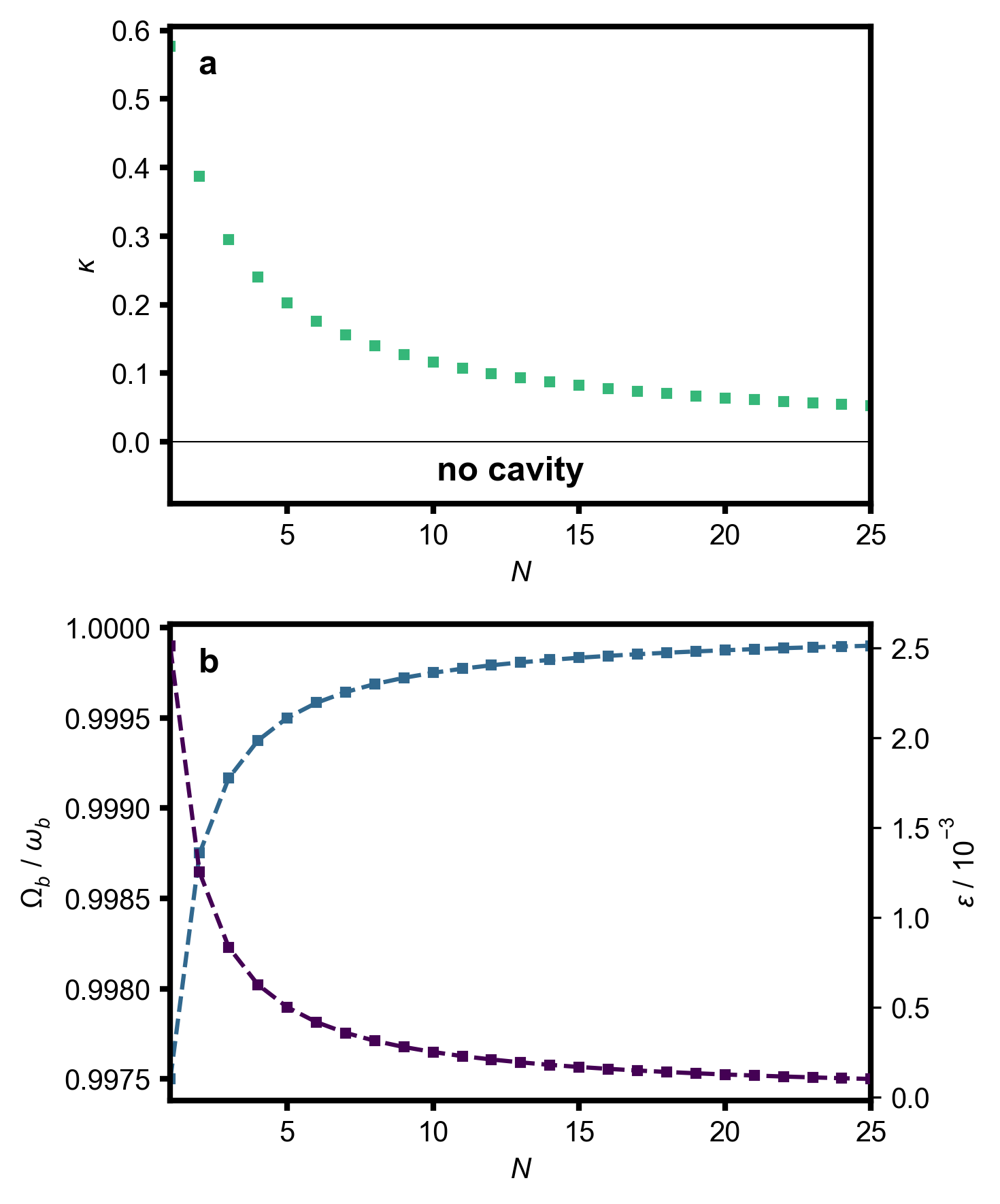}\caption{\label{fig:Nmol}Cavity effects on the PGH model in the collective
regime of $N>1$. (a) $\kappa$, the reaction rate relative to that
from 1D-TST, decreases quickly with increasing $N$. Due to the absence
of molecular bath modes in our model, $\kappa\rightarrow0$ outside
the cavity. With realistic values of $N>20$, we find that $\kappa\rightarrow0$
too; this suggests that the reaction rate approaches the no-cavity
result in the collective regime and cavity effects are lost. (b) Both
the unstable mode barrier frequency $\Omega_{b}$ (blue) and the total
system-polariton coupling $\epsilon$ (purple) approach the no-cavity
limit ($\Omega_{b}/\omega_{b}\rightarrow1$ and $\epsilon\rightarrow0$)
with $N>20$. This is attributable to the tiny single-molecule light-matter
coupling $g$ if we were to consider a realistic experimental set-up
of $N>20$ molecules collectively coupled to the cavity (the polariton
\textquotedblleft large $N$ problem\textquotedblright ). Markers
represent numerical results while dotted lines represent analytical
results obtained from series expansions in $g\sqrt{N-1}$ {[}Eqs.
(\ref{eq:epsilon-analytical}) and (\ref{eq:Omega_b-analytical}){]}.
All plots were generated with the following parameters: equal cavity,
vibrational and barrier frequencies $\omega_{c}=\omega_{v}=\omega_{b}$;
collective light matter coupling $g\sqrt{N}=0.05\omega_{v}$; barrier
height $E_{b}=20k_{\text{B}}T$.}
\end{figure}

The Hamiltonian {[}Eq. (\ref{eq:H-VSC-new}){]} is now in the form
of Eq. (\ref{eq:H_PGH}) and the PGH results may be directly applied.
In principle, $H_{\text{bath}}$ should comprise molecular bath modes
interacting with the polariton and reactive modes through bilinear
couplings drawn from a spectral density. However, single-molecule
analyses of this system suggest that the cavity has the strongest
effect when the molecular bath modes are weakly coupled to the reactive
mode, such that the cavity's coupling is the most prominent among
all the effective bath modes \citep{Lindoy2022,Philbin2022}. We expect
similar results in the presence of $N-1$ non-reacting molecules and
thus consider the zero bath friction limit, i.e. we set $H_{\text{bath}}=0$
and focus on $H_{\text{eff}}$ {[}Eq. (\ref{eq:H_eff}){]}. This simplification
will not change the qualitative outcome. Before we present our numerical
results, we expand, in orders of $g\sqrt{N-1}$, the system-polariton
couplings $g_{\pm}$ (i.e. cavity-induced frictions), the total system-polariton
coupling $\epsilon$ and the unstable mode barrier frequency $\Omega_{b}$,
all of which characterise the cavity's effects on the reaction rate.
In the zero-detuning limit of $\omega_{c}\approx\omega_{v}$, we get
\begin{align}
 &  & g_{\pm} & =\frac{g}{\sqrt{2}}\pm\frac{g^{2}\sqrt{N-1}}{2\sqrt{2}\omega_{v}}+\mathcal{O}\br{g^{3}\br{N-1}},\label{eq:g_pm-analytical}\\
 &  & \epsilon & =\frac{4\omega_{v}^{2}g^{2}}{\br{\omega_{v}^{2}+\omega_{b}^{2}}^{2}}+\mathcal{O}\br{g^{4}\br{N-1}},\label{eq:epsilon-analytical}\\
 & \text{and} & \Omega_{b} & =\omega_{b}-\frac{2\omega_{b}g^{2}}{\omega_{v}^{2}+\omega_{b}^{2}}+\mathcal{O}\br{g^{4}\br{N-1}}\label{eq:Omega_b-analytical}
\end{align}
(see Appendix 2, which uses Ref. \citep{OLeary1990}). The first
term in $g_{\pm}$ represents the single-molecule light-matter coupling
$g$, a first-order process that dominates reaction dynamics and characterises
the cavity-induced friction in single-molecule models \citep{Lindoy2022,Philbin2022}.
The second term represents couplings between the reactive and $N-1$
non-reactive vibrational modes, a second-order process characterised
by $g\br{g\sqrt{N-1}}$. Notice from the expansion coefficients that
light-matter coupling enhances the reaction rate through $\epsilon$
and retards the reaction rate through $\Omega_{b}$, observations
that concur with the single-molecule analysis \citep{Lindoy2022}.
Regardless, to leading order in $g\sqrt{N-1}$, all three parameters
depend only on $g$, the single-molecule coupling, and not on $N$.
For a fixed collective light-matter coupling $g\sqrt{N-1}\approx g\sqrt{N}$
-- the experimentally measurable parameter -- the cavity's effects
(including the friction $g_{\pm}$) diminish as $N$ grows. Such perturbative
results have been observed previously \citep{Poh2022,Kansanen2022,Yang2021}
and is a reminder of the polariton ``large $N$ problem'' \citep{MartinezMartinez2019},
i.e. cavity effects under VSC are mostly characterised by the single-molecule
coupling $g$, a small parameter due to weak photon confinements in
microcavities, and should not be confused with the collective coupling
$g\sqrt{N}$, which is much larger and hence experimentally observable
{[}Fig. (\ref{fig:coupling_orders}){]}.

\begin{figure}
\includegraphics[width=1\columnwidth]{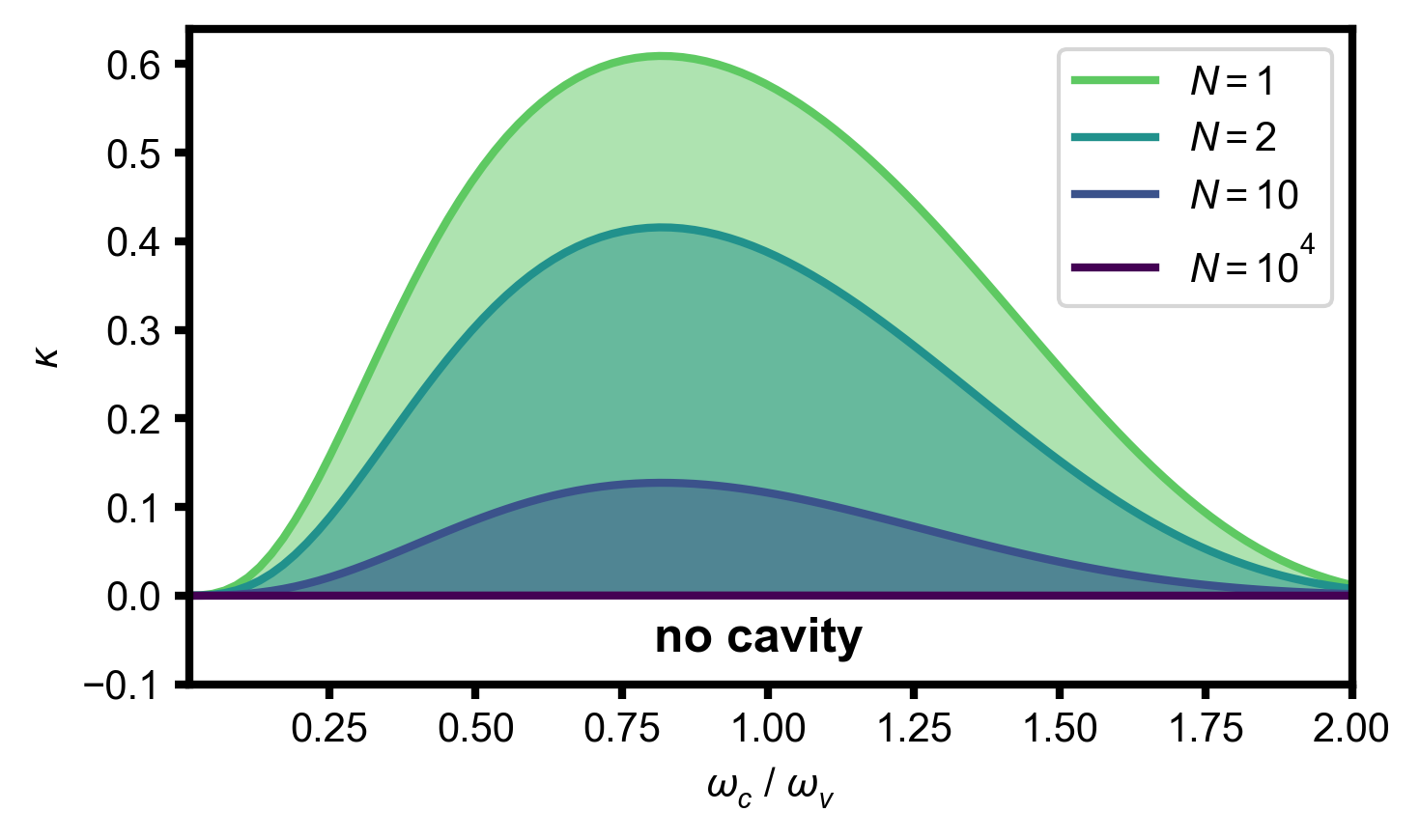}\caption{\label{fig:detuning}Effects of cavity detunings on reaction rates.
In the single-molecule regime ($N=1$), VSC enhances the reaction
rate with maximum modifications observed at slightly negative detunings
($\omega_{c}\lessapprox\omega_{v}$). The same trend is observed in
the collective regime ($N>1$), but with decreasing rate enhancements
at large $N$. Note that $\kappa\rightarrow0$ outside the cavity
(see Fig. (\ref{fig:Nmol}) caption). All plots were generated with
the following parameters: collective light matter coupling $g\sqrt{N}=0.05\omega_{v}$;
barrier height $E_{b}=20k_{\text{B}}T$; barrier frequency $\omega_{b}=\omega_{v}$.}
\end{figure}

The analytical solutions presented in Eqs. (\ref{eq:g_pm-analytical}),
(\ref{eq:epsilon-analytical}) and (\ref{eq:Omega_b-analytical})
agree with numerical simulations conducted at $\omega_{c}=\omega_{v}=\omega_{b}$,
$g\sqrt{N}=0.05\omega_{v}$ and $E_{b}=20k_{\text{B}}T$ {[}Fig. (\ref{fig:Nmol}){]}.
More importantly, $\kappa$, the reaction rate relative to that from
1D-TST, diminishes rapidly with $N$ and approaches the no-cavity
limit after $N>20$ (experimentally, $N\approx10^{6}-10^{12}$ \citep{Pino2015,Daskalakis2017}).
We note that, due to the lack of molecular bath, further removing
the cavity (say, $\omega_{c}\rightarrow0$ or $g\sqrt{N}\rightarrow0$)
would imply $\kappa\rightarrow0$ since no effective bath modes are
present to provide energy for the PGH particle to cross the barrier
(the same conclusion can also be made from Eqs. (\ref{eq:kappa})
and (\ref{eq:DeltaE})). Therefore, any non-zero value of $\kappa$
must be due to the cavity. Next, across different cavity frequencies
$\omega_{c}$ relative to the well vibrational frequency $\omega_{v}$
{[}Fig. (\ref{fig:detuning}){]}, we find maximum rate enhancements
at a value of $\omega_{c}$ slightly below $\omega_{v}$, a result
that agrees with the single-molecule analysis \citep{Lindoy2022}
and has been attributed to the non-linearity of $V\br R$ {[}Eq. (\ref{eq:V(R)}){]}.
Again, such rate enhancements disappear quickly with $N$.

A few comments are now in order:
\begin{enumerate}
\item We note that, by keeping $g\sqrt{N}$ constant at an experimentally-feasible
value, it follows that $g$ must be tiny as $N\rightarrow\infty$.
The importance of this work, however, is not what happens when $N$
is infinitely large, but rather what happens when experimental values
of $N$ is considered, which happens to be large at $\approx10^{6}-10^{12}$
\citep{Pino2015,Daskalakis2017}. Taking this approach allows us to
emphasise the errors incurred by single-molecule models \citep{Sun2022,Lindoy2022,Philbin2022},
which disregard contributions from non-reacting molecules and even
treats the light-matter coupling as the collective one (i.e. incorrectly
set $g\approx g\sqrt{N}$). From another perspective, since $g\sqrt{N}\approx10^{-2}\omega_{v}$,
experimental values of $g$ should be estimated as $\approx\br{10^{-5}-10^{-8}}\omega_{v}$,
a tiny value due to weak photon confinements in microcavities.
\item Even if $g$ were to be kept constant at an \emph{appreciable} value
(which, \emph{we emphasise}, has only been reported in nanoplasmonic
cavities \citep{Chikkaraddy2016,Bitton2022} and does \emph{not} coincide
with experimental observations of rate enhancements), $\kappa$ still
decreases with increasing $N$ since the growing ensemble reduces
the cavity's catalytic efficiency towards the single reacting molecule.
This effect is noteworthy in showing how single-molecule models are
not only limited by their choice of $g$ but also the absence of entropic
effects from an explicit consideration of the ensemble.
\item Note that, unlike Refs. \citep{Galego2019,CamposGonzalezAngulo2020},
the discussion above remains unchanged if the reactive and $N-1$
non-reactive vibrational modes have isotropic spatial alignments.
In that case, the single-molecule light-matter coupling $g=-\boldsymbol{\mu}_{0}'\cdot\boldsymbol{\epsilon}/\sqrt{4\epsilon_{0}\mathcal{V}}$
is replaced by a weighted coupling $-\sqrt{\braket{\mu_{j}^{\prime2}}_{N-1}}/\sqrt{4\epsilon_{0}\mathcal{V}}$,
whereby $\braket{\mu_{j}^{\prime2}}_{N-1}\equiv\br{N-1}^{-1}\sum_{j=1}^{N-1}\br{\boldsymbol{\mu}_{j}'\cdot\boldsymbol{\epsilon}}^{2}$
reflects the average vibrational mode alignment\emph{-squared} and
thus does not vanish in the isotropic limit ($\boldsymbol{\mu}_{j}'$
is the linear change in the dipole moment of molecule $j$). Of course,
the cavity effects still vanish with realistic values of $N$.
\item Also, given how the studied cavity effects depend on $g$ and not
$g\sqrt{N}$ to leading order of the latter term (see Eqs. (\ref{eq:g_pm-analytical}),
(\ref{eq:epsilon-analytical}) and (\ref{eq:Omega_b-analytical})),
it is unlikely that disorder will significantly affect the reaction
rate. Indeed, simulations conducted for $N\le3000$ molecules showed
little rate modifications due to disorder.
\item Interestingly, results from the single molecule model \citep{Lindoy2022}
remained qualitatively unchanged when multiple cavity modes were considered,
with the exception of sharper cavity resonances and weaker cavity
effects \citep{Philbin2022}. As such, our collective VSC analysis
should also hold beyond the single cavity mode limit, provided that
the number of molecules per photon mode remains close to the values
of $N$ here studied \citep{Pino2015a,Daskalakis2017}.
\item Finally, our model and calculations do not show cavity enhancement
effects of friction due to spatial delocalisation of eigenstates \citep{Cao2022}.
We expect these effects to only increase reaction rates by a moderate
amount (as discussed in Supplementary Information S2.3 of Ref. \citep{Du2018}).
Furthermore, these effects will only arise if there are additional
near-field electrostatic interactions among molecules, which have
been ignored in this model in light of how weak they are in the vibrational
regime.
\end{enumerate}

\section*{Conclusion}

In conclusion, thermal reaction rate models, such as the PGH theory,
offer a possible explanation for changes in chemical kinetics within
the single-molecule VSC model \citep{Philbin2022,Lindoy2022,Sun2022}.
Unfortunately, this explanation breaks down with collective VSC commonly
found in vibropolaritonic chemistry experiments. In this regime, $N\approx10^{6}-10^{12}$
molecules simultaneously couple to the cavity \citep{Pino2015,Daskalakis2017}.
As such, the single reacting molecule experiences only a tiny $1/N$
part of the experimentally-observed light-matter interaction, the
remaining of which is shared among the macroscopic number ($N-1$)
of non-reacting molecules, thus negating most rate effects due to
the cavity. Overall, there remains little satisfactory explanation
for rate modifications observed in vibropolaritonic chemistry experiments
and the search continues.

\section*{Acknowledgements}

Acknowledgment is made to the donors of The American Chemical Society
Petroleum Research Fund for support of this research through the ACS
PRF 60968-ND6 Grant. We also thank Arghadip Koner for helpful discussions.

\section*{\label{Appendix 1}Appendix 1: Deriving polariton modes from the
subsystem of cavity and bright modes}

Here, we outline the derivation of the polariton modes (coordinates
$Q_{\pm}$) through a normal mode transformation of the cavity mode
and bright mode (coordinates $q_{c}$ and $Q_{\text{B}}$ respectively).
As an example, we work in the zero cavity detuning limit ($\omega_{c}=\omega_{v}$),
but the same principle applies for any general cavity frequency $\omega_{c}$.
Starting from $H_{\text{nr-c}}$ {[}Eq. (\ref{eq:H_nr-c-bright}){]}
and disregarding the dark modes, we define
\begin{align}
H_{\text{nr-c}}^{\text{eff}} & =H_{\text{nr-c}}-\sum_{\zeta=2}^{N-1}\br{\frac{\dot{Q}_{\zeta}^{2}}{2}+\frac{\omega_{v}^{2}}{2}Q_{\zeta}^{2}}\label{eq:H_nr-c^eff-original}\\
 & =\frac{\dot{Q}_{\text{B}}^{2}}{2}+\frac{\dot{q}_{c}^{2}}{2}+\frac{\omega_{v}^{2}}{2}Q_{\text{B}}^{2}+\frac{1}{2}\br{\omega_{v}q_{c}+2g\sqrt{N-1}Q_{\text{B}}}^{2}\label{eq:H_nr-c^eff}\\
 & \equiv\frac{\dot{Q}_{\text{B}}^{2}}{2}+\frac{\dot{q}_{c}^{2}}{2}+\frac{1}{2}\sum_{k,l=0,1}x_{k}\mathcal{H}_{kl}x_{l},
\end{align}
where $x_{0}=q_{c}$ and $x_{1}=Q_{\text{B}}$, and diagonalise the
Hessian matrix
\begin{align}
\mathcal{H} & =\br{\begin{array}{cc}
\omega_{v}^{2} & 2\omega_{v}g\sqrt{N-1}\\
2\omega_{v}g\sqrt{N-1} & \omega_{v}^{2}+4g^{2}\br{N-1}
\end{array}}
\end{align}
to get the polariton eigenvalues and (normalised) eigenvectors as
\begin{align}
\omega_{\pm}^{2} & =\omega_{v}^{2}+2g^{2}\br{N-1}\pm2g\sqrt{N-1}\sqrt{g^{2}\br{N-1}+\omega_{v}^{2}},\\
Q_{\pm} & =\br{-\frac{g\sqrt{N-1}}{\omega_{v}}\pm\sqrt{1+\frac{g^{2}\br{N-1}}{\omega_{v}^{2}}}}\frac{q_{c}}{K_{\pm}}+\frac{Q_{\text{B}}}{K_{\pm}},\label{eq:Q_pm}
\end{align}
with normalisation constants
\begin{align}
K_{\text{\ensuremath{\pm}}} & =\sqrt{2+\frac{2g^{2}\br{N-1}}{\omega_{v}^{2}}\mp\frac{2g\sqrt{N-1}}{\omega_{v}}\sqrt{1+\frac{g^{2}\br{N-1}}{\omega_{v}^{2}}}}.
\end{align}
Note that the eigenvalues of $\mathcal{H}$ give the square of the
polariton mode frequencies. Also, we have assumed $g\ge0$ without
loss of generality. We may then write $H_{\text{nr-c}}^{\text{eff}}$
{[}Eq. (\ref{eq:H_nr-c^eff}){]} in terms of these polariton normal
modes as 
\begin{align}
H_{\text{nr-c}}^{\text{eff}} & =\sum_{\alpha=\pm}\br{\frac{\dot{Q}_{\alpha}^{2}}{2}+\frac{\omega_{\alpha}^{2}}{2}Q_{\alpha}^{2}}.\label{eq:H_nr-c^eff-new}
\end{align}
Next, noting from Eq. (\ref{eq:Q_pm}) that, after %
\mbox{%
(re-)normalisation%
}, 
\begin{align}
q_{c} & =\br{-\frac{g\sqrt{N-1}}{\omega_{v}}-\sqrt{1+\frac{g^{2}\br{N-1}}{\omega_{v}^{2}}}}\frac{Q_{-}}{K_{-}}\nonumber \\
 & \quad+\br{-\frac{g\sqrt{N-1}}{\omega_{v}}+\sqrt{1+\frac{g^{2}\br{N-1}}{\omega_{v}^{2}}}}\frac{Q_{+}}{K_{+}},\\
Q_{\text{B}} & =\frac{Q_{-}}{K_{-}}+\frac{Q_{+}}{K_{+}},
\end{align}
we write $H_{\text{r-c}}$ {[}Eq. (\ref{eq:H_r-c}){]} in terms of
the polariton modes to get
\begin{align}
H_{\text{r-c}} & =\sum_{\alpha=\pm}\br{2g_{\alpha}\omega_{\alpha}Q_{\alpha}R+2g_{\alpha}^{2}R^{2}},\label{eq:H_r-c-new}
\end{align}
where
\begin{align}
g_{\pm} & =\frac{g}{\omega_{\pm}K_{\pm}}\br{g\sqrt{N-1}\pm\sqrt{\omega_{v}^{2}+g^{2}\br{N-1}}},\label{eq:g_pm}
\end{align}
are the system-polariton couplings and we have noted that $g_{+}^{2}+g_{-}^{2}=g^{2}$.
Combining Eqs. (\ref{eq:H-VSC}), (\ref{eq:H_nr-c^eff-original}),
(\ref{eq:H_nr-c^eff-new}) and (\ref{eq:H_r-c-new}) and completing
the squares give Eqs. (\ref{eq:H-VSC-new}) and (\ref{eq:H_eff}).

\section*{\label{Appendix 2}Appendix 2: Series expansions of $g_{\pm}$, $\epsilon$
and $\Omega_{b}$ }

Here, we outline the approach taken to expand $g_{\pm}$, $\epsilon$
and $\Omega_{b}$ in orders of $g\sqrt{N-1}$. For simplicity, we
set the cavity detuning as zero, i.e. $\omega_{c}=\omega_{v}$. Then,
$g_{\pm}$ may be expanded directly from Eq. (\ref{eq:g_pm}) to get
Eq. (\ref{eq:g_pm-analytical}). Next, we find the unstable and stable
modes by performing a normal mode transformation on the Hamiltonian
$H_{\text{eff}}$ {[}Eq. (\ref{eq:H_eff}){]} near the barrier region.
Noting that $V\br R\approx-\frac{1}{2}\omega_{b}^{2}R^{2}+E_{b}$
in this region {[}Eq. (\ref{eq:V(R)}){]}, we define
\begin{align}
H_{\text{eff}}^{\br b} & =\frac{\dot{R}}{2}+\sum_{\alpha=\pm}\frac{\dot{Q}_{\alpha}^{2}}{2}+E_{b}+\frac{1}{2}\sum_{k,l=0,1,2}y_{k}K_{kl}^{\br b}y_{l},
\end{align}
with $y_{0}=R$, $y_{1}=Q_{-}$ and $y_{2}=Q_{+}$, and diagonalise
the force constant matrix
\begin{align}
K^{\br b} & =\left(\begin{array}{ccc}
-\omega_{b}^{2}+4g_{-}^{2}+4g_{+}^{2} & 2g_{-}\omega_{-} & 2g_{+}\omega_{+}\\
2g_{-}\omega_{-} & \omega_{-}^{2} & 0\\
2g_{+}\omega_{+} & 0 & \omega_{+}^{2}
\end{array}\right)\label{eq:K^b}
\end{align}
to get an unstable coordinate $u$ with negative eigenvalue $-\Omega_{b}^{2}$
($\Omega_{b}\in\mathbb{R}^{+}$) and two stable coordinates $\curbr{s_{k}}$
with positive eigenvalues $\curbr{\Omega_{k}^{2}}$ ($\Omega_{k}\in\mathbb{R}^{+}$).
Since $K^{\br b}$ is an arrowhead matrix, the secular equation and
modal matrix (which are used to find the eigenvalues and eigenvectors)
have simple forms \citep{OLeary1990}. In particular, $-\Omega_{b}^{2}$
satisfy the following equation
\begin{align}
\Omega_{b}^{2} & =\omega_{b}^{2}\br{1+\frac{4g_{-}^{2}}{\omega_{-}^{2}+\Omega_{b}^{2}}+\frac{4g_{+}^{2}}{\omega_{+}^{2}+\Omega_{b}^{2}}}^{-1},
\end{align}
which is solved using perturbation theory in orders of $g\sqrt{N-1}$
to get Eq. (\ref{eq:Omega_b-analytical}). Also, $c_{00}$, the contribution
of $u$ to $R$, can be found from the modal matrix as
\begin{align}
c_{00} & =\br{\sqrt{1+\br{\frac{2g_{-}\omega_{-}}{-\Omega_{b}^{2}-\omega_{-}^{2}}}^{2}+\br{\frac{2g_{+}\omega_{+}}{-\Omega_{b}^{2}-\omega_{+}^{2}}}^{2}}}^{-1},
\end{align}
which, using Eq. (\ref{eq:Omega_b-analytical}) and noting the definition
of $\epsilon$ in Eq. (\ref{eq:epsilon}), gives Eq. (\ref{eq:epsilon-analytical})
through a series expansion in $g\sqrt{N-1}$.

\end{document}